# Solution Processed CMOS compatible Carbon Nano-dots Based Heterojunction for Enhanced UV Detector

R. Maiti[1], S. Mukherjee[2], T. Dey[3] and S. K. Ray[1,4*]

[1]Department of Physics, Indian Institute of Technology Kharagpur, Kharagpur-721302, India.

[2]Advanced Technology Development Center, Indian Institute of Technology, Kharagpur-721302, India.

[3]School of Nanoscience & Nanotechnology, Indian Institute of Technology, Kharagpur-721302, India

[4]S. N. Bose National Centre for Basic Sciences, Kolkata-700106, India

*physkr@phy.iitkgp.ac.in

Carbon nanostructures technology has recently emerged as a key enabler for next-generation optoelectronic devices including deep UV detectors and light sources which is promising in health and environment monitoring. Here, we report the fabrication of solution processed Carbon nano-dots (CNDs)/n-Si heterojunction showing broadband spectral response with a peak responsivity of ~ 1.25 A/W in UV (~300 nm) wavelength. The surface topography and chemical information of synthesized CNDs via a facile synthesis route have been characterized showing the presence of surface chemical states resulting broad optical emission. The CNDs/n-Si photo diodes exhibit very low dark current (~500 pA), excellent rectification ratio (~$5\times10^3$), and very good photo-modulation in UV region. Given the solution-processing capability of the devices and extraordinary optical properties of CNDs, the use of CNDs will open up unique opportunities for future high-performance, low-cost DUV photo detectors.

Ultraviolet (UV) photodetectors have drawn extensive attention owing to their wide range of applications, such as water purification, flame detection, ozone layer monitoring, and optical communications[1, 2] etc. Traditional UV detectors are fabricated using thin films of a wide band gap semiconductor, such as ZnO, GaN, and AlGaN etc[3, 4]. However, the growth of epitaxial films with cost intensive fabrication processes is challenging with Si-CMOS technology. Silicon is known to be a good choice for detectors in the visible and near infrared (VIS-NIR) wavelength range[5-8] but not so useful for UV detection. Solar-blind DUV PDs for extremely harsh environments are still at their early stage of development.

On the other hand, carbon based nanostructures have gained lot of interests recently for various applications[9], with graphene, being one of the most studied materials over the last decade[10]. But, the use of graphene for optoelectronic devices is limited due to its zero-band gap. Recently, carbon nanodots (CNDs)[11] have attracted much attention owing to their several unique properties, such as, excellent photostability, quantum confinement, biocompatibility, size tunable photoluminescence (PL) and exceptional multi-photon excitation[12,13] characteristics. Though CNDs appear potentially attractive for applications in optoelectronics[14], energy storage devices[15], sensors[16], and biomedical imaging[17], their integration over large area on mature planer CMOS platform is yet to be explored.

Here, we report the fabrication and characteristics of Si-CMOS compatible CNDs heterojunctions, exhibiting very low dark current (~500 pA) and excellent rectification (~$5\times10^3$) behavior with a high on-off ratio (~$10^4$) at reverse bias. Fabricated p-CNDs/n-Si heterojunctions have shown high photo-responsivity and detectivity in the UV region (300 nm) making it useful for realizing UV photodetectors with enhanced performances.

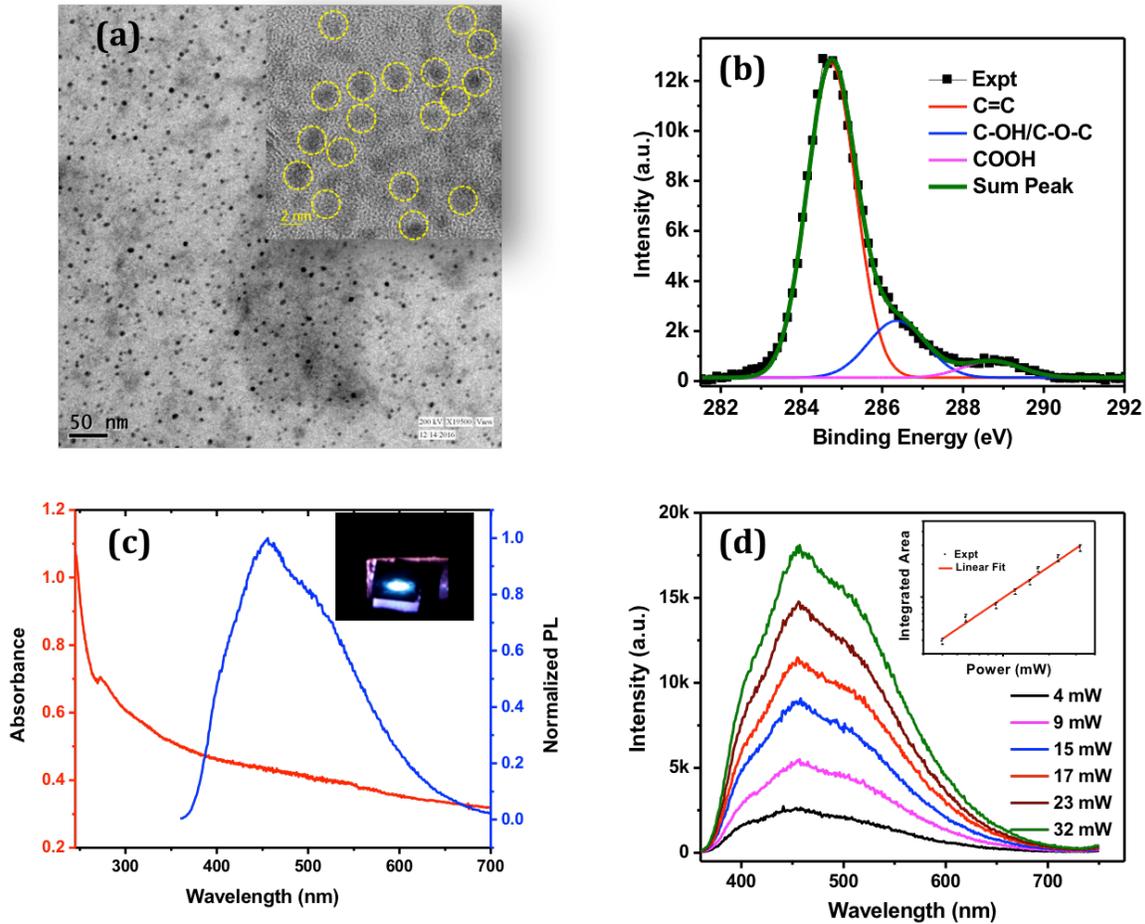

FIG. 1. Physical characterization of as synthesized CNDs (a) TEM image of prepared CNDs showing spherical in shape. HRTEM micrograph of synthesized CNDs (inset). Average diameter of CNDs found to be ~2 nm. Typical (b) C1s XPS spectrum showing presence of different surface oxidation states, (b) Absorption spectrum with strong UV absorption showing the potential of demonstrating UV detector & broad PL spectrum excited by 325 nm He-Ne laser (inset- Pl emission photograph). (d) PL spectrum of as synthesized CNDs sample with different excitation power showing linear variation (inset).

The detailed synthesis process of CNDs has been reported elsewhere[18]. Briefly, CND was synthesized via solvothermal treatment of orange juice in ethanol, followed by washing and gradient centrifugation[19]. The result is presented in Fig. 1(a), which shows the formation of spherical nano-dots. The average size of the CNDs is found to be ~2 nm from the HRTEM micrograph shown in the inset of Fig. 1(a). X-ray photo electron spectra (XPS) (PHI 5000 Versa Probe II, INC, Japan) showing the binding energy of C1s electrons of CNDs (Fig. 1(b)) which can be fitted by Gaussian function as C=C or C-C bonds at 284.6 eV, epoxy (C-O-) or hydroxyl (C-OH) bonds at 286.4 eV and carboxyl (-COOH) bonds at 288.8 eV, in agreement with previous reports[20]. The optical absorption spectrum of as-synthesized CNDs displays a strong peak at 270 nm due to the π-π* transition of aromatic >C=C< bonds and extends up to 800 nm without any other significant features where extinction coefficient of CNDs was found to be ~45 mL mg$^{-1}$ mm$^{-1}$ at 300 nm (figure 1(c)). We acquire the steady state photoluminescence spectra of drop-casted CNDs using a He-Cd laser of wavelength 325 nm as a function excitation source and a TRIAX-320 monochromator fitted with Hamamatsu R928 photomultiplier detector. The broad PL spectra for as-deposited sample is arising due to the combination of the sp$^2$ hybridized core state and sp$^3$ hybridized surface state emission[19] as shown in figure 1 (c). We have carried out excitation power dependent photoluminescence (PL) measurement for as synthesized CNDs by varying the incident power from 4 mW to 32 mW, as shown in Fig. 1(d). The power dependent integrated intensity plot shows linear variation with the exponent value of 0.95 for different excitation power (inset).

As-synthesized CNDs, which are p-type in nature due to presence of electron withdrawing oxygen functional groups on the surface [21], were spin coated on n-Si substrates to fabricate p-n heterojunction diode (Fig. 2(a)) with a corresponding cross-sectional SEM image of the fabricated device (figure 2b). For electrical contacts, thin Au dots (device area ~0.2 mm$^2$) were deposited as top electrodes by thermal evaporation on the CNDs. Large area Al was also deposited for achieving an Ohmic back contact on n-Si, followed by vacuum annealing at 200°C for 5 min. To get an understanding on the role playing by CNDs, a control sample was also fabricated without CNDs layer on n-Si. Fig. 2(c) shows a semi-log scale plot of the current-voltage (*I-V*) characteristics of CND/n-Si hetero-junction under dark and illuminated conditions studied using semiconductor parameter analyzer (4200-SCS). The dark current is found to be extremely low and saturates at around ~1 × 10$^{-9}$A at -2 V bias, resulting in a photo-to-dark current ratio of nearly $10^4$ at -1 V. The asymmetric natures of the I-V curves indicate the formation of the hetero-junction between n-Si and p-CNDs.

The diode equation $I = I_0 \left( e^{\frac{eV_A}{\eta k_B T}} - 1 \right)$ has been used to fit the *I-V* characteristic for a small positive voltage zone ($\eta$ is the ideality factor, $k_B$ is the Boltzmann constant, $I_0$ is the dark saturation current, $V_A$ is the applied bias voltage, $T$ is the room temperature). The ideality factor ($\eta$) of this photodiode is estimated to be 1.9, as shown in Fig. 2(d). The relatively high value of ideality factor may be attributed to the surface defects formed in CND/n-Si interface and the high density of trap-states, which are inherent in the CNDs due to the colloidal-synthesis route[5, 21]. The typical rectification behavior of fabricated heterojunction diode depicting half wave rectification for 75 Hz ac input signal is shown in Fig. 2(e). The rectification ratio (the ratio of forward bias current to the reverse bias current at the same bias voltage) of this hetero-junction is found to be ~$10^4$ at 2 V.

The fabricated CND/n-Si heterojunction exhibits fast switching behavior when illuminating with a 325 nm He-Cd laser source. Fig. 3(a) shows the variation of photocurrent $(I_{ph})$ for different laser powers ($P_{opt}$) at a fixed bias voltage of -3 V. With increase of the incident laser power from 0.5 mW to 3 mW, the current ON/OFF ratio is seen to increase steadily due to the enhancement of photocurrent. The spectral photocurrent response was measured by standard lock-in method using a broadband light source (xenon arc lamp). The responsivity (*R*) of a photodetector can be defined as the ratio of photocurrent density ($J_{ph}$) to the power of incident light ($P_{opt}$) corresponding to a particular wavelength incident on the detector. The spectral responsivity (for different bias voltages) of our Au/CND/n-Si photodetector is shown in Fig. 3(b). The peak responsivity occurs at 300 nm, and is observed to be 1.25, 0.77 and 0.50 A/W for a bias voltage -4V, -2V and -1V, respectively. This may be attributed to the presence of strong absorption peak of CNDs in the UV wavelength range as shown in Fig. 1(b).

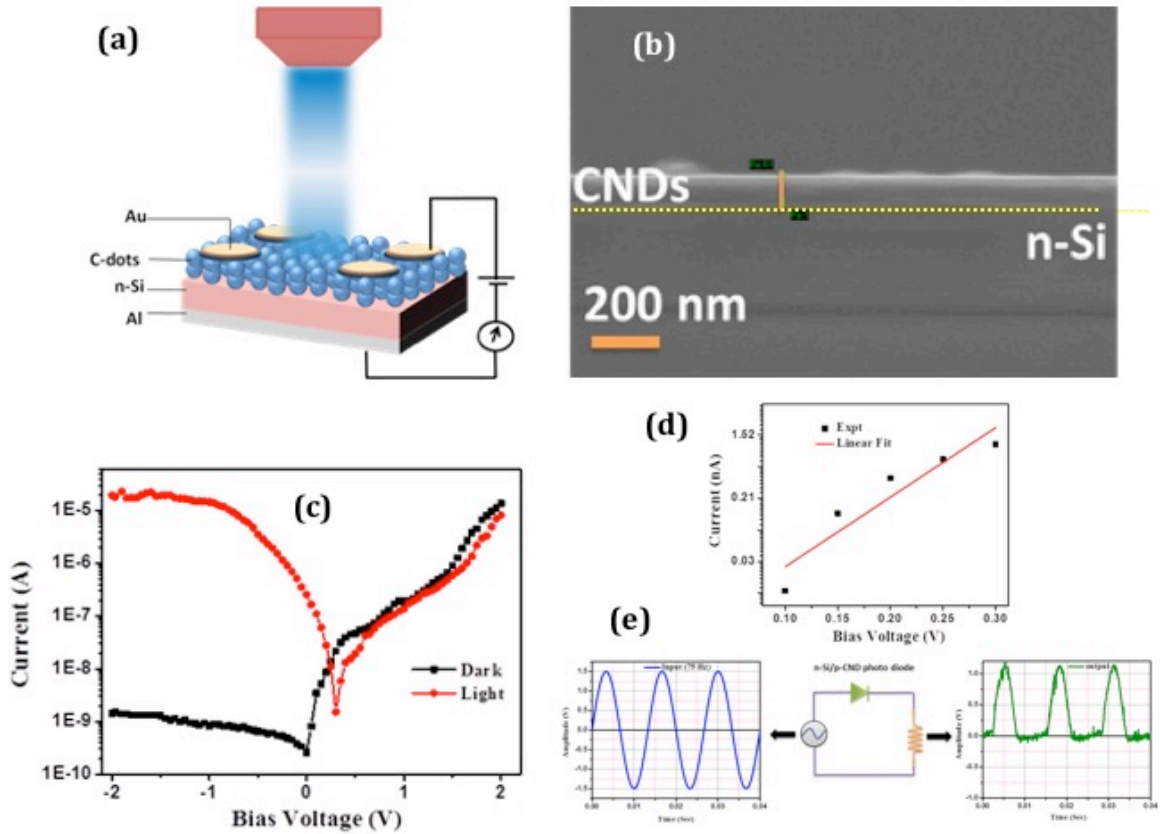

FIG.2.Electrical characteristics of heterojunction, (a) Device schematic of the proposed Au/CND/n-Si UV photodetector. (b) The cross-sectional view of fabricated device using FESEM. (c) Typical I-V curves of the device (plotted on semi-logarithmic scale) measured under the dark and light illumination. (d) Fitted I-V curve to estimate ideality factor of the diode. (e) The typical rectification characteristics of p-CND/n-Si diode for 75 Hz of ac input signal.

The spectral responsivity also shows a small hump at 1000 nm due to the absorption of Silicon substrate, shown in the inset of Fig. 3(b). A photodetector's detectivity can be approximated as[7], $D(\lambda) = \frac{R(\lambda)}{\sqrt{2qJ_d}}$, where $q$ is the electronic charge, $R(\lambda)$ is the responsivity, $J_d$ is the photocurrent density. The detectivity value for the CND heterojunction device is found to be highest at 300 nm, and is 2.06 x$10^{14}$, 1.63 x$10^{14}$, 1.37 x$10^{14}$ cm.Hz$^{0.5}$/W at a bias voltages of -4V, -2V and -1V, respectively, as shown in Fig. 3(c). Different figure of merits of CND based photodetectors reported in the literature are summarized in Table I, showing superior performances in terms of spectral responsivity and detectivity mainly due to the strong UV absorption shown by CNDs despite being solution processed device fabrication.

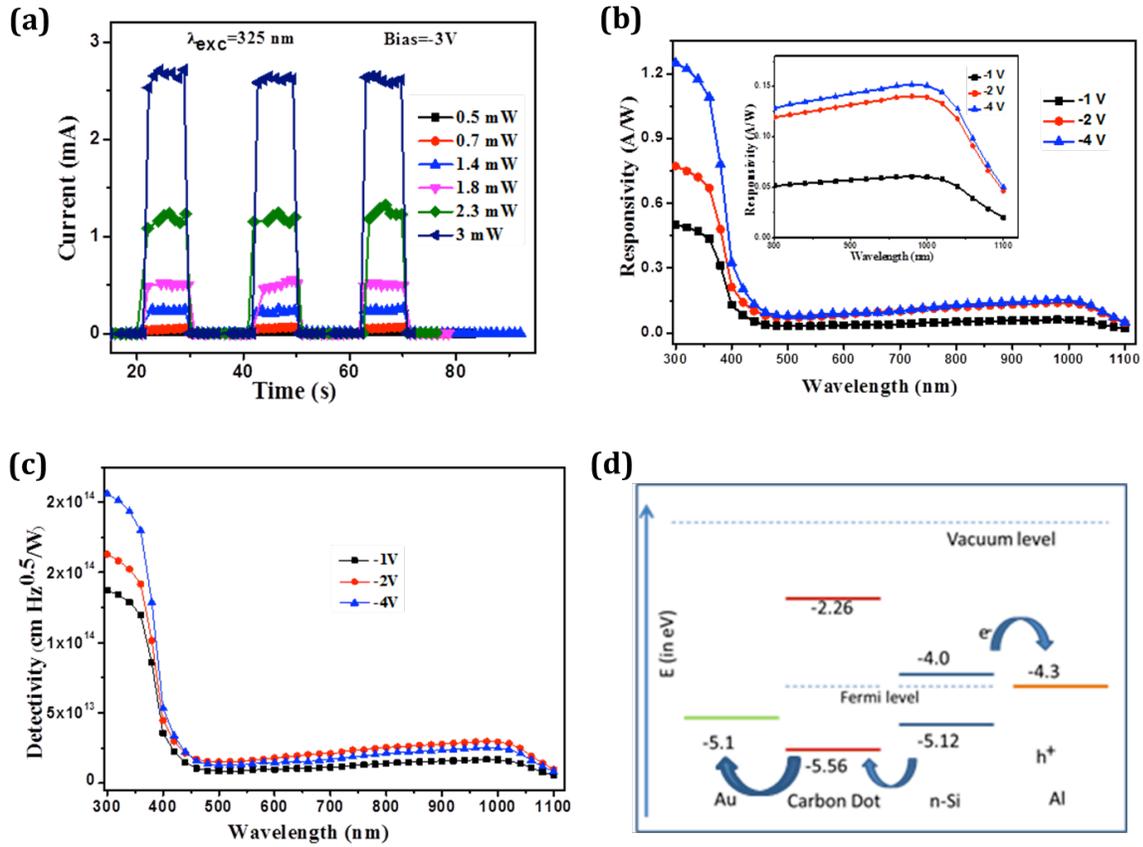

FIG.3.Typical switching characteristics of the heterojunction device with fixed bias voltage (-3 V) with different illuminated powers under 325 nm laser excitation. (b) Spectral responsivity of CND based UV-photodetector under different bias voltages. (inset: responsivity zoomed near 1000 nm). (c) Spectral detectivity of the UV detector for different bias voltages. (d) Schematic energy level diagram for the fabricated CND heterojunction diode.

To explain the observed photo-diode behavior in detail, a schematic energy band diagram of the CND/n-Si heterojunction device is depicted in Fig. 3(d) with previously reported material parameters from the literature[6]. In thermal equilibrium, a built-in electric field exists in the depletion region due to the band offset resulting in sweeping of the generated carriers when illuminated which gives rise to a small photocurrent. With the increase of reverse bias, higher values of responsivity lead to the efficient collection of photocarriers at the metal electrodes. This enhanced photoresponse in UV region indicates that, solution processed CNDs based devices are attractive for future flexible and transparent nanoscale optoelectronic devices.

Table I: Comparison of different parameters for reported carbon nanoparticle based photodetectors

| Material | Device type | Responsivity | Spectral range | Detectivity | Reference |
|---|---|---|---|---|---|
| Carbon Quantum Dots (CQDs) | 2- terminal p-n junction Si NW/CQD core shell heterojunction | 0.353 A/W | 400-1100nm | $\sim 10^9$ | 22 |
| Graphene Quantum Dots (GQDs) | 2-terminal MSM diode Ag/GQD/Au | 2.1mA/W | DUV (<300 nm) | $\sim 10^{12}$ | 23 |
| n-doped Graphene Quantum Dots (n-GQDs) | 2-terminal MSM diode on IDE Au/n-GQD/Au | 325 V/W | UV-NIR | -- | 24 |
| Graphene Quantum Dots (GQDs) | 2-terminal tunneling diode Graphene/GQD/Graphene | 0.2-0.5 A/W | 300-1050nm | $>10^{11}$ | 25 |
| Carbon Nano dots (CNDs) | 2- terminal p-n junction n-Si/p-CNDs heterojunction | 0.5 A/W | 300-1100 nm | $\sim 10^{14}$ | **Our present work** |

In summary, we demonstrate a CND/n-Si heterojunction broadband photodiode, which shows enhanced UV sensitivity. The synthesized CNDs by facile chemical route from orange juice display strong UV absorption with a broadband PL emission. The CNDs based photodiode exhibits superior performance in terms of very low dark current density, high rectification ratio, stable switching behavior and high responsivity at UV region. The unavailability of the deep UV source limits our measurement up to 300 nm. This work opens up great opportunities of using CNDs in high-performance, low-cost UV photodetectors, compatible with planar CMOS technology.

**Acknowledgements:**

R. Maiti and S. Mukherjee acknowledge financial assistantship from MHRD, India. T. Dey acknowledges DST for providing INSPIRE fellowship (IF-160592). XPS facility at IIT Kharagpur under the DST FIST project is gratefully acknowledged.